\documentclass[onecolumn]{svjour3}
\journalname{Nonlinear Dynamics}

\usepackage[margin=1in]{geometry}
\usepackage{graphicx}
\usepackage{mathptmx}
\usepackage{latexsym,amssymb,amsmath}
\usepackage{float}
\usepackage{hyperref}

\title{Trajectory-free approximation of phase space structures using the trajectory divergence rate}
\author{Gary K. Nave Jr. \and
	Peter J. Nolan \and
	Shane D. Ross}
\institute{G.K. Nave (Corresponding author) \and P.J. Nolan \and S.D. Ross\at 
	Engineering Mechanics Program\\
	Virginia Polytechnic Institute and State University\\
	\email{gknave@vt.edu} \\
	Tel: +1-804-397-0700}
\date{Received: date / Accepted: date}

\begin{document}
\maketitle
\begin{abstract}
	This paper introduces the trajectory divergence rate, a scalar field which locally gives the instantaneous attraction or repulsion of adjacent trajectories. This scalar field may be used to find highly attracting or repelling invariant manifolds, such as slow manifolds, to rapidly approximating hyperbolic Lagrangian coherent structures, or to provide the local stability of invariant manifolds. This work presents the derivation of the trajectory divergence rate and the related trajectory divergence ratio for 2-dimensional systems, investigates their properties, shows their application to several example systems, and presents their extension to higher dimensions.
\end{abstract}

\section*{Acknowledgments}
This work was supported by National Science Foundations grants 1520825, 1537349, and 1821145 and by the Biological Transport (BioTrans) Interdisciplinary Graduate Education Program at Virginia Tech.

\section{Introduction}

To better understand the properties of mathematical models and experimental measurements, it is often convenient to look at the geometric structure of the flow of a resulting vector field. There often exist lower-dimensional manifolds which dominate the attraction and repulsion, swirling, or shearing of trajectories advecting under the flow. Methods to find such structures have been applied to better understand topics including plant pathogen spread \cite{schmale2015highways}, animal locomotion \cite{nave2018global,peng2008theupstream}, seabird foraging patterns \cite{kai2009top}, geophysical flows \cite{lekien2005pollution,wiggins2005dynamical}, chemical reactions \cite{zhong2017tube}, comet distributions \cite{dellnitz2005transport}, and structural mechanics \cite{wiggins2001impenetrable}.

In autonomous systems, the simplest geometric structure of interest is the fixed point, which is a 0-dimensional invariant manifold in the flow. Stable, unstable, and center manifolds of a fixed point may be calculated through a number of classical methods, including ``growing'' the stable or unstable manifolds by integrating the eigendirections of fixed points backward or forward in time \cite{koon2008dynamical}. However, in the context of weak stable submanifolds, these methods begin to break down \cite{nave2018global}. Some weak submanifolds are part of a class of geometric structures known as \textit{slow manifolds} which exhibit a separation of time scales \cite{kuehn2016multiple}, attracting or repelling other trajectories in phase space. This difference in time scales between motion along a slow manifold and the motion normal to it allows them to be classified as normally hyperbolic invariant manifolds (NHIMs) \cite{wiggins2013normally}.

Geometric structure may also be present in the absence of fixed points or slow manifolds. Recent developments in dynamical systems have led to several useful generalizations of some key geometric features. Distinguished hyperbolic trajectories generalize stable and unstable manifolds to aperiodic flows, and are identified using the ``$M$-function'' \cite{madrid2009distinguished}. Hyperbolic coherent structures represent dynamically evolving transport barriers in flows \cite{shadden2011lagrangian,shadden2005definition}. Although the methods of coherent structures are typically situated within the context of fluid dynamics, such structures have applications to the flow of general vector fields \cite{aldridge2006direct,gawlik2009lagrangian,nave2018global,tanaka2010mathematical}. Detecting and analyzing the underlying structures of flows gives a better understanding of how the system evolves, whether that flow represents the motion of a fluid or some other general dynamical system.

Methods to identify coherent structures may be based on integrated trajectory information or may be calculated from the instantaneous vector field for the entire volume. Most state of the art methods are trajectory-based, using finite-time integration of trajectories to calculate coherent structures \cite{balasuriya2018generalized,shadden2011lagrangian}. There is a wide variety of trajectory-based methods to identify coherent structures or coherent sets, including transfer operator methods \cite{dellnitz2001algorithms,froyland2015rough}, topological methods \cite{allshouse2012detecting,budivsic2015finite}, and stretching-based methods such as the finite-time Lyapunov exponent (FTLE) \cite{shadden2005definition}; see Hadjighasem et al. \cite{hadjighasem2017critical} for a review. 
Lagrangian descriptors, similarly, calculate properties of the flow along trajectories and can be used to detect ``distinguished trajectories'' \cite{lopesino2017theoretical,madrid2009distinguished}. However, trajectory-based methods involve significant computational resources, requiring trajectory integration over an ensemble of initial conditions \cite{ameli2014development,brunton2010fast}.

There is much to be gained by looking at the instantaneous information of vector fields. Although the trajectory-dependent coherent structures are more robust to the flow, the short time behavior of these structures may be of interest \cite{haller2010localized}. Vector field schemes are also much more computationally efficient, and their changes can be tracked in time for nonautonomous flows. Historically, most vector field-based methods have focused on elliptic, or vortex-like, coherent structures \cite{chakraborty2005relationships}. More recent work has developed the notion of objective Eulerian coherent structures for 2-dimensional flows \cite{serra_objective_2016}, which include hyperbolic and parabolic structures in addition to objectively defined elliptic coherent structures. However, while objectivity is necessary for detecting, for instance, vortex-like coherent structures in a fluid, objectivity may be a disadvantage in other examples of dynamical systems \cite{haller_variational_2011,lopesino2017theoretical}.

\subsection{Main result}
This paper introduces the \textit{trajectory divergence rate} for 2-dimensional vector fields, given by,
\begin{equation*}
\dot{\rho} = \mathbf{n}^\dagger\mathbf{S}\mathbf{n},
\end{equation*}
where $\mathbf{n} = \mathbf{R}\mathbf{v}/\left|\mathbf{v}\right|$ is the unit normal vector field, with $R=\left(\begin{smallmatrix}
0 & -1 \\ 1 & 0
\end{smallmatrix}\right)$ giving a $90^\circ$ rotation, and $\mathbf{S}=\tfrac{1}{2}\left(\nabla\mathbf{v}+(\nabla\mathbf{v})^\dagger\right)$ is the rate-of-strain tensor, representing the symmetric component of the Jacobian of the system. In this work, the dagger, \((\cdot)^\dagger\), indicates the matrix transpose to avoid confusion with time-integrated methods. The trajectory divergence rate is an inherent property of $C^1$ vector fields, measuring the extent to which the trajectory passing through each point instantaneously repels or attracts nearby trajectories. This paper will show that this quantity may be used as a diagnostic tool to approximate slow manifolds and hyperbolic coherent structures by showing regions of strong instantaneous repulsion or attraction.

Instantaneous attraction and repulsion has been considered previously, through other metrics such as the normal infinitesimal Lyapunov exponent (NILE) \cite{haller2010localized} and the strain acceleration tensor \cite{haller2001lagrangian}. These methods have primarily been applied to partition the space or look into regions of local stability or instability, particularly within applications of turbulence. The trajectory divergence rate introduced herein is intended to serve as a ``rough and ready'' method for approximating hyperbolic, or stretching-based, geometric structures in the flows of general nonlinear dynamical systems. Much like local curvature \cite{desroches2011canards}, this quantity may be thought of as an inherent property of continuously differentiable vector fields, showing the instantaneous local divergence or convergence of nearby trajectories. Under certain conditions, the regions of highest local divergence or convergence serve to approximate finite-time coherent structures.

The idea of stability is asymptotic in nature; a stable invariant manifold is one for which nearby trajectories stay close for $t\to\infty$. Although transport barriers and invariant manifolds in flows are calculated based on the long-term dynamics of the system, and therefore the long-term repulsion of the manifold in question, the instantaneous repulsion of invariant manifolds provides additional insights into the character of an invariant manifold, as regions of a globally attracting invariant manifold may be instantaneously repelling \cite{haller2001lagrangian,tallapragada2017globally}. The trajectory divergence rate is easily computable, requiring only the vector field and its gradient, and can serve as a useful diagnostic in the search for influential geometric structures in flows.

Section \ref{s:Background} gives the mathematical preliminaries and notation to provide the mathematical context for the divergence rate. Section \ref{s:Derivation} shows the derivation of the trajectory divergence rate and discusses its properties. Section \ref{s:Application} shows several applications of the divergence rate over a different situations in which it may prove useful. Section \ref{s:HigherDimension} extends the trajectory divergence rate from 2-dimensional to higher dimensional systems and provides a 3-dimensional example. Finally, Section \ref{s:Summary} provides some discussion about the work of this paper to conclude the work.

\section{Background and notation}\label{s:Background}
To begin, consider a general 2-dimensional, autonomous ordinary differential equation
\begin{equation}
\mathbf{\dot x} = \mathbf{v}(\mathbf{x}), \quad \mathbf{x} \in U \subseteq \mathbb{R}^2, \, t\in \mathbb{R},
\end{equation}
with its time-$T$ mapping all initial conditions forward to their positions after a duration $T$,
\begin{equation}\label{eq:flowmap}
\begin{split}
& \mathbf{F}_T : U \rightarrow U, \quad T \in \mathbb{R}, \\
& \mathbf{x}_0 \mapsto \mathbf{x}_T = \mathbf{x}(T;\mathbf{x}_0).
\end{split}
\end{equation}
% Similarly, 
For any $\mathbf{x} \in U$ with $\mathbf{v}(\mathbf{x}) \ne \mathbf{0}$ (that is, excluding equilibrium points), one can define the following unit vector fields parallel and normal to the governing vector field $\mathbf{v}(\mathbf{x})$, respectively,
\begin{equation}
\begin{split}
& \mathbf{e}(\mathbf{x}) = \frac{\mathbf{v}(\mathbf{x})}{|\mathbf{v}(\mathbf{x})|}, \\
& \mathbf{n}(\mathbf{x}) = \mathbf{R} \mathbf{e}(\mathbf{x})
, \quad   
\mathbf{R} = \left(
\begin{array}{cc}
0 & -1 \\
1 & ~ 0 
\end{array}
\right).
\end{split}
\label{unit_vector_fields}
\end{equation}
The gradient of the time-$T$ flow map $\nabla\mathbf{F}_T$ defines a mapping from vectors based at $\mathbf{x}_0$, such as $\mathbf{e}(\mathbf{x}_0)$ and $\mathbf{n}(\mathbf{x}_0)$, to vectors based at $\mathbf{x}_T$, showing how those vectors deform with the flow. The tangent vector, in general, maps to the new tangent direction, but the normal vector does not map to the normal direction at time $T$ due to the shear of the flow.

\subsection{Trajectory-normal repulsion rate}
\begin{figure}
\centering
\includegraphics[width=3in]{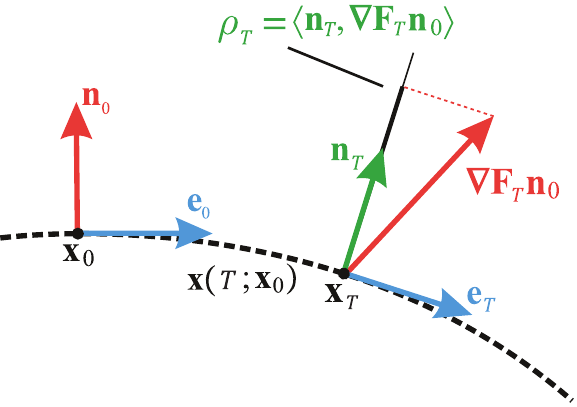}
\caption{Geometry of the trajectory-normal repulsion rate, reproduced from \cite{nave2018global}.}
\label{fig:normal-repulsion-factor-2D}
\end{figure}

For any trajectory passing through a point $\mathbf{x}_0 \in U$ with 
$\mathbf{v}(\mathbf{x}_0) \ne \mathbf{0}$, the {\it trajectory-normal repulsion rate} $\rho_T(\mathbf{x}_0)$ \cite{haller_variational_2011} over the time interval $[0, T]$ may be defined locally as the projection of $\nabla\mathbf{F}_T (\mathbf{x}_0) \mathbf{n}_0 $ onto the new normal vector
$\mathbf{n}_T$,
\begin{equation}
\rho_T(\mathbf{x}_0) = \langle  \mathbf{n}_T , \nabla \mathbf{F}_T (\mathbf{x}_0) \mathbf{n}_0 \rangle,
\label{eq:reprate}
\end{equation}
where $\mathbf{n}_T = \mathbf{n}(\mathbf{x}_T) = \mathbf{n}(\mathbf{F}_T (\mathbf{x}_0))$ and $\langle\cdot,\cdot\rangle$ is the usual inner product in $\mathbb{R}^2$. 
As illustrated in Figure \ref{fig:normal-repulsion-factor-2D}, $\rho_T(\mathbf{x}_0)$ is a measure of the growth of infinitesimal perturbations normal to the invariant manifold containing $\mathbf{x}_0$ over the time interval $[0,T]$.
If the projection $\rho_T(\mathbf{x}_0) >1$, then infinitesimal perturbations normal to the trajectory through $\mathbf{x}_0$ grow over the time interval $[0,T]$. Note that, although overall growth over the duration $T$ may be repelling (attracting), it is possible for the invariant manifold to be instantaneously attracting (repelling) \cite{tallapragada2017globally}.

This scalar field $\rho_T(\mathbf{x}_0)$ can be used to extract the most influential invariant manifolds in the flow, in the sense that it reveals those manifolds that normally repel (or attract) other manifolds at the largest rate. Using this trajectory-normal repulsion rate, one can calculate, for example, slow-manifolds, such as those found in the examples below.

\subsection{Trajectory-normal repulsion ratio}
A related quantity is the {\it trajectory-normal repulsion ratio} \cite{haller_variational_2011}, which is the ratio of normal repulsion to the tangential stretching along an invariant manifold passing through the point $\textbf{x}_0$ over the time interval $[0, T]$,
\begin{equation}
\nu_T(\mathbf{x}_0) = \frac{\rho_T(\mathbf{x}_0)}{\left|\nabla\mathbf{F}_T (\mathbf{x}_0) \mathbf{e}_0\right|}.
\end{equation}
Where the trajectory-normal repulsion ratio \(\nu_T(\mathbf{x}_0)>1\), the normal stretching dominates the tangential stretching of the curve. 

Both the trajectory-normal repulsion rate and trajectory-normal repulsion ratio can be written in terms of the right Cauchy-Green tensor $\mathbf{C}_T (\mathbf{x}_0)$, well-known from continuum mechanics \cite{truesdell2004non}, as well as its use in FTLE and LCS calculations \cite{haller_variational_2011},
\begin{equation}
\begin{aligned}
\rho_T(\mathbf{x}_0) 
& = \sqrt{ \frac{
\left| \mathbf{v}(\mathbf{x}_0)\right|^2 {\rm det} \mathbf{C}_T (\mathbf{x}_0)}{
\langle \mathbf{v}(\mathbf{x}_0),\mathbf{C}_T (\mathbf{x}_0) \mathbf{v}(\mathbf{x}_0) \rangle
} } \\
\nu_T(\mathbf{x}_0) 
& =  \frac{
\left| \mathbf{v}(\mathbf{x}_0)\right|^2 \sqrt{{\rm det} \mathbf{C}_T (\mathbf{x}_0)}}{
\langle \mathbf{v}(\mathbf{x}_0),\mathbf{C}_T (\mathbf{x}_0) \mathbf{v}(\mathbf{x}_0) \rangle
}
\end{aligned}
\label{eq:rhoT}
\end{equation}
As a matter of notation, in this work \((\cdot)_T\) will indicate a value calculated over the interval $[0, T]$. Note that $T$ may be positive or negative. Because of their dependence on the normal vector in the derivation of these expressions, these scalar fields both remain defined only for 2-dimensional systems.

When both $\rho_T(\mathbf{x})>1$ and $\nu_T(\mathbf{x})>1$ for all $\mathbf{x}\in\gamma$, where $\gamma$ is an invariant manifold, and $\gamma$ is a ridge of the $\rho_T$-field, $\gamma$ is a constrained Lagrangian coherent structure \cite{haller_variational_2011}, in the sense that the variational search for attracting or repelling curves is constrained to the space of invariant manifolds.

\section{The trajectory divergence rate}\label{s:Derivation}
The trajectory-normal repulsion rate \(\rho_T\) may be useful in finding attracting (or repelling) structures in a 2-dimensional flow \cite{haller_variational_2011}, but calculation of the time-$T$ flow map over the domain of interest is computationally expensive. Therefore, this work seeks an instantaneous measure that gives the leading order behavior of this scalar field.

\begin{figure*}
\centering
\includegraphics[width=4.5in]{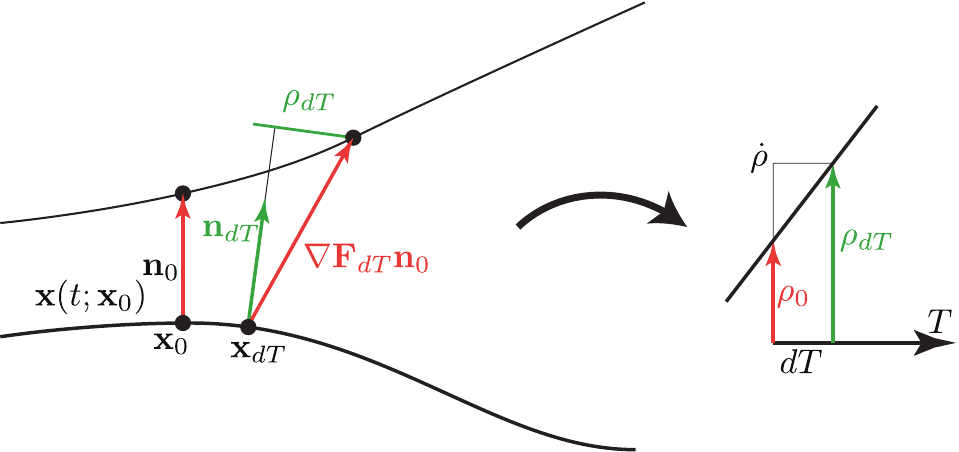}
\caption{This schematic shows the relationship between the divergence rate and the time-$T$ repulsion rate $\rho_T$. The divergence rate is the instantaneous rate of change of the repulsion rate, which is therefore signified as $\dot{\rho}$.}
\label{fig:div-rate-schematic}
\end{figure*}

For scalar and tensor fields, the dependence on $\mathbf{x}_0$ will be notationally dropped for clarity, as it will be understood. For small time \(T\), the right Cauchy-Green tensor, \(\mathbf{C}_T  \), may be expanded in terms of integration time $T$,
\begin{equation}
\mathbf{C}_T   = \mathbf{C}_0+\left.\frac{d\mathbf{C}_T}{dT}\right|_{T=0}T + \frac{1}{2} \left.\frac{d^2\mathbf{C}_T}{dT^2}\right|_{T=0}T^2+ \cdots
\label{eq:CG Expansion}	
\end{equation}
Because all derivatives are evaluated at $T=0$, $\left.\tfrac{d}{dt}\right|_{t=0}=\left.\tfrac{d}{dT}\right|_{t=0}$. The derivatives of the right Cauchy-Green tensor are given by the Rivlin-Ericksen tensors \cite{truesdell2004non},
\begin{equation}
\begin{aligned}
\frac{d^k\mathbf{C}_T}{dt^k} &= \nabla\frac{d\mathbf{x}}{dt}+\left(\nabla\frac{d\mathbf{x}}{dt}\right)^\dagger, &k = 1,\\
\frac{d^k\mathbf{C}_T}{dt^k} &= \nabla\frac{d^k\mathbf{x}}{dt^k} + \left(\nabla\frac{d^k\mathbf{x}}{dt^k}\right)^\dagger & \\
\hfill& + \sum_{i=1}^{k-1}\left(\!\begin{array}{c}
k \\ i
\end{array}\!\right)\left(\nabla\frac{d^i\mathbf{x}}{dt^i}\right)^\dagger \nabla\frac{d^{k-i}\mathbf{x}}{dt^{k-i}}, &k \geq 2.
\end{aligned}
\end{equation}
For small $T\ll1$, the leading order behavior is given by the first Rivlin-Ericksen tensor $(\nabla\mathbf{v}+(\nabla\mathbf{v})^\dagger)$. Neglecting higher order dependence on $T$, the expansion of the Cauchy-Green tensor (\ref{eq:CG Expansion}) simplifies to,
\begin{equation}
\mathbf{C}_T   = \mathbf{I} + 2 \mathbf{S} T
+ \mathcal{O}(T^2),
\label{eq:Expansion2}
\end{equation}
where $\mathbf{S}$ represents the symmetric rate-of-strain tensor, defined as,
\begin{equation}
\mathbf{S}   = \tfrac{1}{2} \left( \nabla \mathbf{v}  + \nabla \mathbf{v} ^\dagger \right).
\end{equation}
Note that in $\mathbb{R}^2$, we will denote the eigenvalues of $\mathbf{S}$ as $s_1,s_2$, with $s_1\leq s_2$.

The expansion of the Cauchy-Green tensor in (\ref{eq:CG Expansion}) makes it possible to perform a Taylor expansion of the trajectory-normal repulsion rate, $\rho_T$ from (\ref{eq:reprate}) and Fig. \ref{fig:normal-repulsion-factor-2D} for a small integration time $T\ll1$. For a 2-dimensional system, the following identity allows the expansion of the determinant within the trajectory-normal repulsion rate,
\begin{equation}
{\rm det} (\mathbf{A} + \mathbf{B}) = {\rm det} \mathbf{A} + {\rm det} \mathbf{B} + {\rm det} \mathbf{A} \cdot {\rm tr}(\mathbf{A}^{-1} \mathbf{B}), 
\end{equation}
Neglecting the higher order dependence on $T$ in (\ref{eq:Expansion2}) admits the substitution \(\mathbf{C}_T   = \mathbf{I} +2T\mathbf{S} \). Therefore, \(\det \mathbf{C}_T \), can be expressed as,
\begin{equation}\label{eq:detC}
\begin{aligned}
\det \mathbf{C}_T &= \det(\mathbf{I} +2T\mathbf{S}) \\
&= \det(\mathbf{I} ) + \det(2T\mathbf{S}) + \det(\mathbf{I}) \text{tr} (2T\mathbf{I} ^{-1}\mathbf{S}) \\
&= 1 + 4T^2\det(\mathbf{S}) + 2T\text{tr}(\mathbf{S}) \\
&= 1 + 2T\text{tr}(\mathbf{S}) + \mathcal{O}(T^2).
\end{aligned}
\end{equation}

To finish the expansion of  (\ref{eq:rhoT}), the same substitution \(\mathbf{C}_T   = \mathbf{I} +2T\mathbf{S} \) gives the following result.
\begin{equation}\label{eq:expand}
\begin{aligned}
\frac{\left|\mathbf{v}\right|^2}{\mathbf{v}^\dagger \mathbf{C}_T\mathbf{v}} &= \frac{\left|\mathbf{v}\right|^2}{\left|\mathbf{v}\right|^2+2T \mathbf{v}^\dagger\mathbf{S}\mathbf{v}+\mathcal{O}(T^2)} \\
&= \frac{1}{1+\frac{1}{\left|\mathbf{v}\right|^2}2T\mathbf{v}^\dagger\mathbf{S}\mathbf{v}+\mathcal{O}(T^2)} \\
&= 1-2T \frac{\mathbf{v}^\dagger \mathbf{S}\mathbf{v}}{\left|\mathbf{v}\right|^2} + \mathcal{O}(T^2)
\end{aligned}
\end{equation}

Combining these two substitutions gives the following relation for the trajectory-normal repulsion ratio.
\begin{equation}
\begin{aligned}
\rho_T  &= \sqrt{\frac{
| \mathbf{v} |^2 \det\mathbf{C}_T }{
\mathbf{v} ^\dagger\mathbf{C}_T  \mathbf{v} 
}} \\
&= \sqrt{\left(1 + 2T\text{tr}\left(\mathbf{S}\right)+\mathcal{O}(T^2)\right)\left(1-2T \frac{\mathbf{v} ^\dagger\mathbf{S} \mathbf{v} }{\left|\mathbf{v} \right|^2}+\mathcal{O}(T^2)\right)} \\
&= 1+\left(\text{tr}(\mathbf{S} )-\frac{\mathbf{v} ^\dagger \mathbf{S} \mathbf{v} }{\left|\mathbf{v} \right|^2}\right)T + \mathcal{O}(T^2)
\end{aligned}
\end{equation}
Neglecting higher order terms for small $T$, 
\begin{equation}
\rho_T  = 1+\left(\text{tr}(\mathbf{S} )-\frac{\mathbf{v} ^\dagger \mathbf{S} \mathbf{v} }{\left|\mathbf{v} \right|^2}\right)T
\label{eq:local Rho}
\end{equation}
Therefore, the leading order behavior of \(\rho_T \) for small $T$ is given entirely by the quantity $\dot \rho = \left.\frac{d \rho_T}{dT}\right|_{T=0}$,
\begin{equation}
\dot \rho   = \text{tr}(\mathbf{S} )-\frac{\mathbf{v} ^\dagger \mathbf{S} \mathbf{v} }{\left|\mathbf{v} \right|^2}
\label{eq:Leading Order Behavior}
\end{equation}
which is the {\it trajectory divergence rate}. Fig. \ref{fig:div-rate-schematic} shows a schematic of the geometric interpretation of this derivation.

This quantity is independent of the choice of the time parameter \(T\), and, furthermore, does not require integration to be calculated. It is dependent solely on the given vector field $\mathbf{v}(\mathbf{x})$	and its gradient through the rate-of-strain tensor
$\mathbf{S} $. As shown in Appendix \ref{ap: normal derivation}, for 2-dimensional systems, this expression reduces to simply
\begin{equation}
\dot{\rho}  = \mathbf{n}^\dagger\mathbf{S} \mathbf{n}
\label{eq:DivRate}
\end{equation}

The instantaneous rate of normal repulsion is given by a quadratic form on the  rate-of-strain tensor by the unit normal vector. The trajectory divergence rate can also be derived via the following expression for the rate of change of length of an infinitesimal vector $\ell$,
\begin{equation}
\frac{1}{2}\frac{d}{dt}\left|\ell\right|^2 = \ell^\dagger\mathbf{S}\ell.
\end{equation}

\subsection{Trajectory divergence ratio}
Following the same procedure as the expansion of the repulsion rate, the trajectory-normal repulsion ratio may be expanded by,
\begin{equation}
\nu_T  
= \frac{
\left| \mathbf{v} \right|^2 \sqrt{{\rm det} \mathbf{C}_T  }}{
\langle \mathbf{v} ,\mathbf{C}_T   \mathbf{v}  \rangle
},
\label{eq:nuT}
\end{equation}
for small \(T\) to find its instantaneous rate of growth. From (\ref{eq:detC}),
\begin{equation}
\sqrt{\det \mathbf{C}_T } = 1 + T\text{tr}(\mathbf{S} )+\mathcal{O}(T^2),
\end{equation}
and using (\ref{eq:expand}),
\begin{equation}
\begin{aligned}
\nu_T  & = \left(1 + T\text{tr}(\mathbf{S} )+\mathcal{O}(T^2)\right)\left(1-2T \frac{\mathbf{v} ^\dagger \mathbf{S} \mathbf{v} }{\left|\mathbf{v} \right|^2}+\mathcal{O}(T^2)\right) \\
& = 1 + T\left(\text{tr}(\mathbf{S} )-2\frac{\mathbf{v} ^\dagger \mathbf{S} \mathbf{v} }{\left|\mathbf{v} \right|^2}\right)+\mathcal{O}(T^2)
\end{aligned}
\end{equation}
And the rate of \(\nu_T\) is given as 
\begin{equation}
\dot{\nu} = \frac{\mathbf{v}^\dagger\left(\text{tr}(\mathbf{S})\mathbf{I}-2 \mathbf{S}\right)\mathbf{v}}{\left|\mathbf{v}\right|^2}
\end{equation}
Similar to the trajectory divergence rate $\dot{\rho}$, the trajectory divergence ratio is dependent only on the rate-of-strain tensor and therefore does not require the calculation of trajectories. These scalar fields give a measurement of the instantaneous stretching of normal vectors throughout phase space and can be used to find the most attracting and repelling structures with much less computational cost.

\subsection{Physical interpretation of the trajectory divergence rate}
\begin{figure*}
\centering
\includegraphics[height=1.5in]{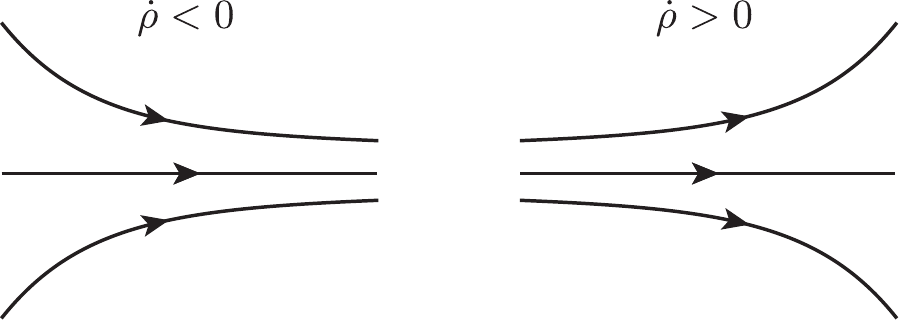}
\caption{Schematic of the physical interpretation of the trajectory divergence rate $\dot{\rho}$. Negative values indicate converging trajectories while positive values indicate diverging trajectories.}
\label{fig:DivRateSchema}
\end{figure*}

The trajectory divergence rate provides a scalar measurement of how much a trajectory is attracting or repelling nearby trajectories, representing the time-normalized slope of the normal distance between nearby trajectories, as visualized in Figure \ref{fig:div-rate-schematic}. Therefore, as shown in Figure \ref{fig:DivRateSchema}, a positive divergence rate indicates diverging trajectories, a negative divergence rate indicates converging trajectories, and a zero divergence rate shows the regions of the flow where trajectories are parallel. To visualize this, consider the simple linear saddle flow.

\begin{figure}
\centering
\begin{minipage}{0.45\textwidth}
\centering
\includegraphics[height=2.7in]{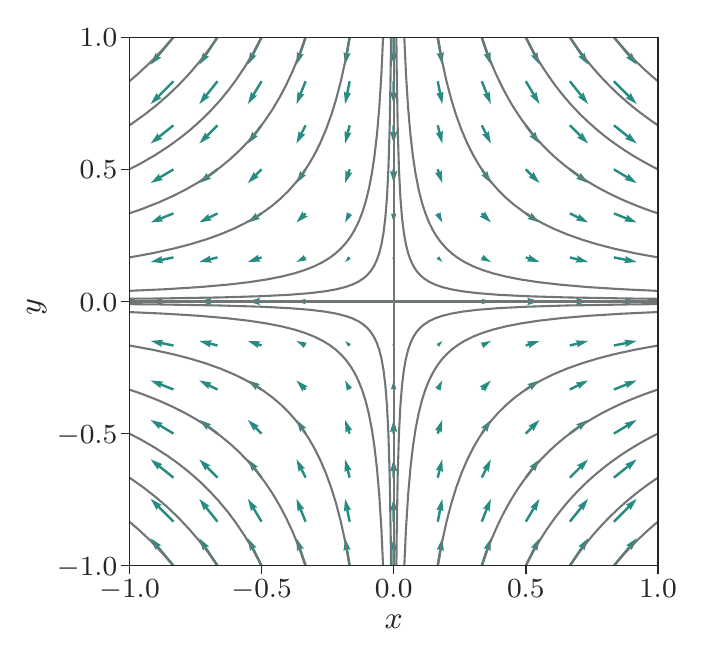}
\end{minipage}
\begin{minipage}{0.45\textwidth}
\centering
\includegraphics[height=2.8in]{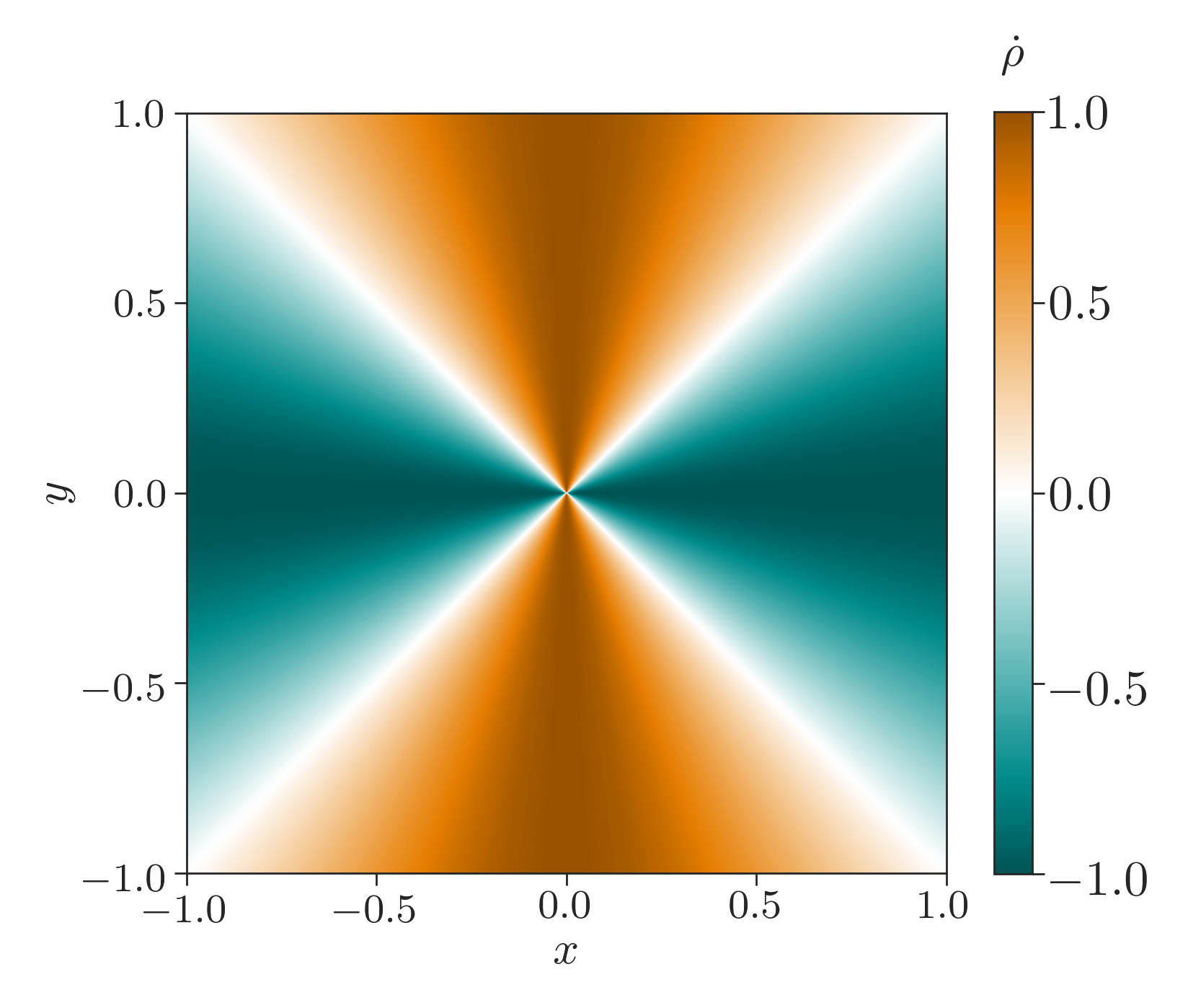}%RepRate-saddle
\end{minipage}
\caption{The phase portrait (top) and trajectory divergence rate field (bottom) for Example \ref{ex:saddle}.}
\label{fig:saddle}
\end{figure}

\example{\it\textendash \label{ex:saddle} Linear saddle flow.} \\
The saddle-point flow represents the simplest case of stable and unstable manifolds. The system is given by
\begin{equation}
\begin{aligned}
\dot{x} & = x, \\
\dot{y} & = -y.
\end{aligned}
\end{equation}
As is visible in Figure \ref{fig:saddle}, the linear saddle flow repels trajectories from the $y$-axis and attracts them to the $x$-axis in forward time.

The unit normal vector and rate-of-strain tensor are given by 
\begin{equation}
\mathbf{n} = \frac{1}{\sqrt{x^2+y^2}}\left[\begin{array}{c}
y \\
x
\end{array}\right], \quad
\mathbf{S} = \left[\begin{array}{cc}
1 & 0 \\
0 & -1
\end{array}\right].
\end{equation}
From these, the trajectory divergence rate is computed to be,
\begin{equation}
\dot{\rho} = \frac{y^2-x^2}{x^2+y^2}.
\label{eq:SaddleDivRate}
\end{equation}

Figure \ref{fig:saddle} shows trajectories in phase space and the trajectory divergence rate of the linear saddle. From (\ref{eq:SaddleDivRate}) and the trajectory divergence rate in the figure, trajectories are converging when $x^2>y^2$ and diverging when $y^2>x^2$. Trajectories are parallel where $y=\pm x$, as shown in white. The ridges and troughs of the trajectory divergence rate field give the most repelling and attracting curves in the field: the vertical and horizontal axes, respectively. Interestingly, the forward and backward finite-time Lyapunov exponents are both uniform for the linear saddle flow, indicating no structure \cite{haller_variational_2011}.

\subsection{Remarks on the trajectory divergence rate}

\subsubsection{(Lack of) objectivity of the trajectory divergence rate}
The trajectory divergence rate (\ref{eq:DivRate}) is not an objective quantity, as a scalar quantity such as $\dot{\rho}$ would be objective only if it remained unchanged under any translation and rotation of reference frame \cite{truesdell2004non,serra_objective_2016}. In other words, objective scalar values remain invariant under transformations belonging to the Special Euclidean group $SE(n)$. Because the trajectory divergence rate depends on the tangent vectors, which are not objective, the quantity itself is not objective. However, as shown in the context of Lagrangian descriptors, objectivity is not always a desirable trait \cite{lopesino2017theoretical}. For the example of a rotating saddle flow, the finite-time Lyapunov exponent, which is an objective quantity, gives no structure at all. However, in that example, Lagrangian descriptors, which are not objective, are able to show the rotating saddle at every snapshot in time. Under situations where objectivity is necessary, the trajectory divergence rate may not be the appropriate tool to use. However, objectivity is not always a desirable property, and makes no sense in general abstract phase spaces.

\subsubsection{Relationship to Objective Eulerian Coherent Structures}
In their paper introducing objective Eulerian coherent structures \cite{serra_objective_2016}, Serra and Haller introduce two objective quantities to calculate these structures: the stretching rate $\dot{p}$ and shear rate $\dot{q}$. These equations depend on the tangent vectors $x'$ of a general curve $\gamma$ parametrized by its arc length $s$.
\begin{equation}\label{eq:stretch-rate}
\dot{p} = \frac{\langle x'(s),\mathbf{S}x'(s)\rangle}{\langle x'(s),x'(s)\rangle}
\end{equation}
\begin{equation}\label{eq:shear-rate}
\dot{q} = \frac{\langle x'(s),\left(\mathbf{S}\mathbf{R}-\mathbf{R}\mathbf{S}\right)x'(s)\rangle}{\langle x'(s),x'(s)\rangle}
\end{equation}
These scalar functions are objective because they depend generally on $\mathbf{S}$, which is an objective tensor, and tangent vectors to a curve $\gamma$, which is not dependent on the vector field. If these curves are restricted to trajectories following the vector field rather than general curves, then their tangent vectors become the vector field $x'=\mathbf{v}$, and $\dot{p}$ and $\dot{q}$ become quadratic forms on vectors $\mathbf{v}$ like the trajectory divergence rate and trajectory divergence ratio. However, they lose the objectivity that is central to the previous work. As discussed above, there are situations where objectivity is less important, so trajectory-based variations of the stretch rate $\dot{p}_v$ and shear rate $\dot{q}_v$ which depend on the vector field may prove useful. Considering the unit tangent vector $\mathbf{e}=\mathbf{v}/\left|\mathbf{v}\right|$, these are given by,
\begin{equation}
\begin{aligned}
\dot{p}_v &= \langle\mathbf{e}, \mathbf{S}\mathbf{e}\rangle, \\
\dot{q}_v &= \langle\mathbf{e}, \left(\mathbf{S}\mathbf{R}-\mathbf{R}\mathbf{S}\right)\mathbf{e}\rangle.
\end{aligned}
\end{equation}
Together with the trajectory divergence rate introduced above, these three quadratic forms measure the instantaneous rates of tangential stretching, normal stretching, and shear of the vector field. The trajectory stretch rate $\dot{p}_v$ and trajectory shear rate $\dot{q}_v$ are worth further exploration in future studies.

	\subsubsection{Normal hyperbolicity of trajectories}
	On the other hand, removing the restriction of the trajectory divergence rate to trajectories of the vector field to instead calculate the normal repulsion of a general surface, the trajectory divergence rate becomes an objective scalar value just like the stretching and shear rates above. In fact,the expression for the trajectory divergence rate in Eq. (\ref{eq:DivRate}) is similar to an existing quantity has been applied to normal vectors of candidate Lagrangian coherent structure as a test of hyperbolicity, using gradients of the finite-time Lyapunov exponent field to determine the tangent and normal directions \cite{green2010using}. The trajectory divergence rate, in contrast, uses the vector normal to trajectories of the underlying dynamical system. From this observation, it is clear that the trajectory divergence rate gives the normal hyperbolicity field of the vector field. Because of this, the trajectory divergence rate may be a useful metric for finding normally hyperbolic structures in a flow. This idea is explored further in Sec. \ref{ss:Application-SlowMnflds}.

\section{Applications of the trajectory divergence rate}\label{s:Application}
As a measure of normal attraction and repulsion of trajectories of a system, the trajectory divergence rate can be applied to a variety of special cases to identify influential structures in dynamical systems. It may serve as a good approximation for hyperbolic Lagrangian coherent structures or as a method to identify slow manifolds. Additionally, it may be relevant to calculate the normal hyperbolicity of a particular trajectory for applications in control.

\subsection{Approximation of slow manifolds and normally hyperbolic invariant manifolds}\label{ss:Application-SlowMnflds}
Given the interpretation of the trajectory divergence rate as a measure of normal hyperbolicity, it becomes a natural tool to identify normally hyperbolic invariant manifolds (NHIMs). One of the key examples of NHIMs is in the study of slow manifolds of multiple time scale systems \cite{kuehn2016multiple}. In such systems there is a lower-dimensional manifold on which most of the dynamics occur, referred to as the slow manifold. This is typically conceptualized as an attracting manifold, but may be repelling in some cases. Outside of the slow manifold, the motion moves more quickly onto (or away from) the slow manifold. %The slow manifold is instantaneously attracting or repelling in the fast direction.

\begin{figure}
\begin{minipage}{3.2in}
\centering
\includegraphics[width=1.8in]{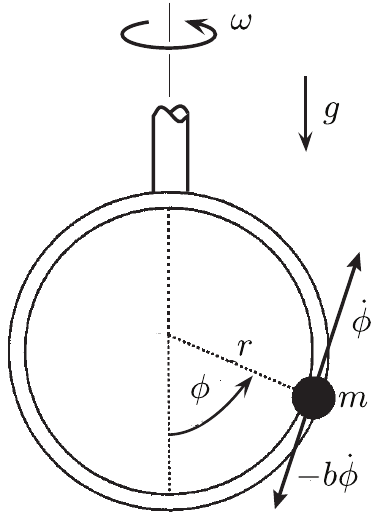}
\end{minipage}
\begin{minipage}{3.2in}
\centering
\includegraphics[width=3.2in]{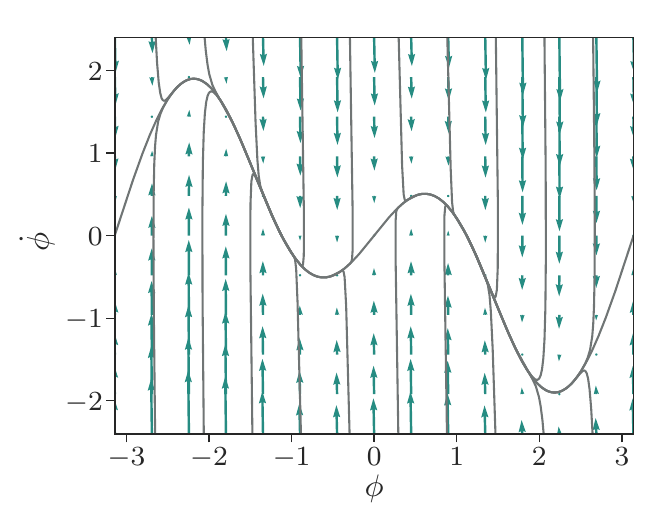}
\end{minipage}
\caption{\label{fig:rotHoopA} (Left) Schematic of an overdamped bead on a rotating hoop, Coulomb friction directly opposes the motion along the hoop, while the entire hoop rotates about the vertical axis with rotational velocity $\omega$. (Right) The phase portrait for the model of Example \ref{ex:rotHoop} defined by (\ref{eq:rotHoop}), using $\varepsilon = 0.02$ and $\mu = 2.3$.}
\end{figure}

\example{\it\textendash Overdamped bead on a rotating hoop\label{ex:rotHoop}.} \\
This example comes from Strogatz \cite[Section 3.5]{strogatz_nonlinear_2014}, providing a nice example of a slow-fast system. The system conceptualizes a bead moving along a circular hoop of radius $r$ while the hoop is spinning with constant angular velocity $\omega$ about a vertical axis. Considering a dimensionless time $T=\tfrac{b}{mg}$ and parameters $\mu=\tfrac{r\omega^2}{g}$ and $\varepsilon=\tfrac{m^2gr}{b^2}$, the forces on the body reduce to the system, 

\begin{equation}
\begin{aligned}
\dot{\phi} & = \Omega, \\
\dot{\Omega } & = \frac{1}{\varepsilon}\left(\sin\phi(\mu\cos\phi - 1) - \Omega\right).
\end{aligned}
\label{eq:rotHoop}
\end{equation}

When the damping coefficient $b$ is large, the parameter $\varepsilon$ becomes very small, and trajectories collapse quickly to the curve $\Omega = \sin\phi (\mu \cos\phi - 1)$ due to the high damping in the system, and then ooze along it toward one of several fixed points. This system is illustrated in Fig. \ref{fig:rotHoopA}.

\begin{figure}
\begin{minipage}{3.2in}
\centering
\includegraphics[width=3.2in]{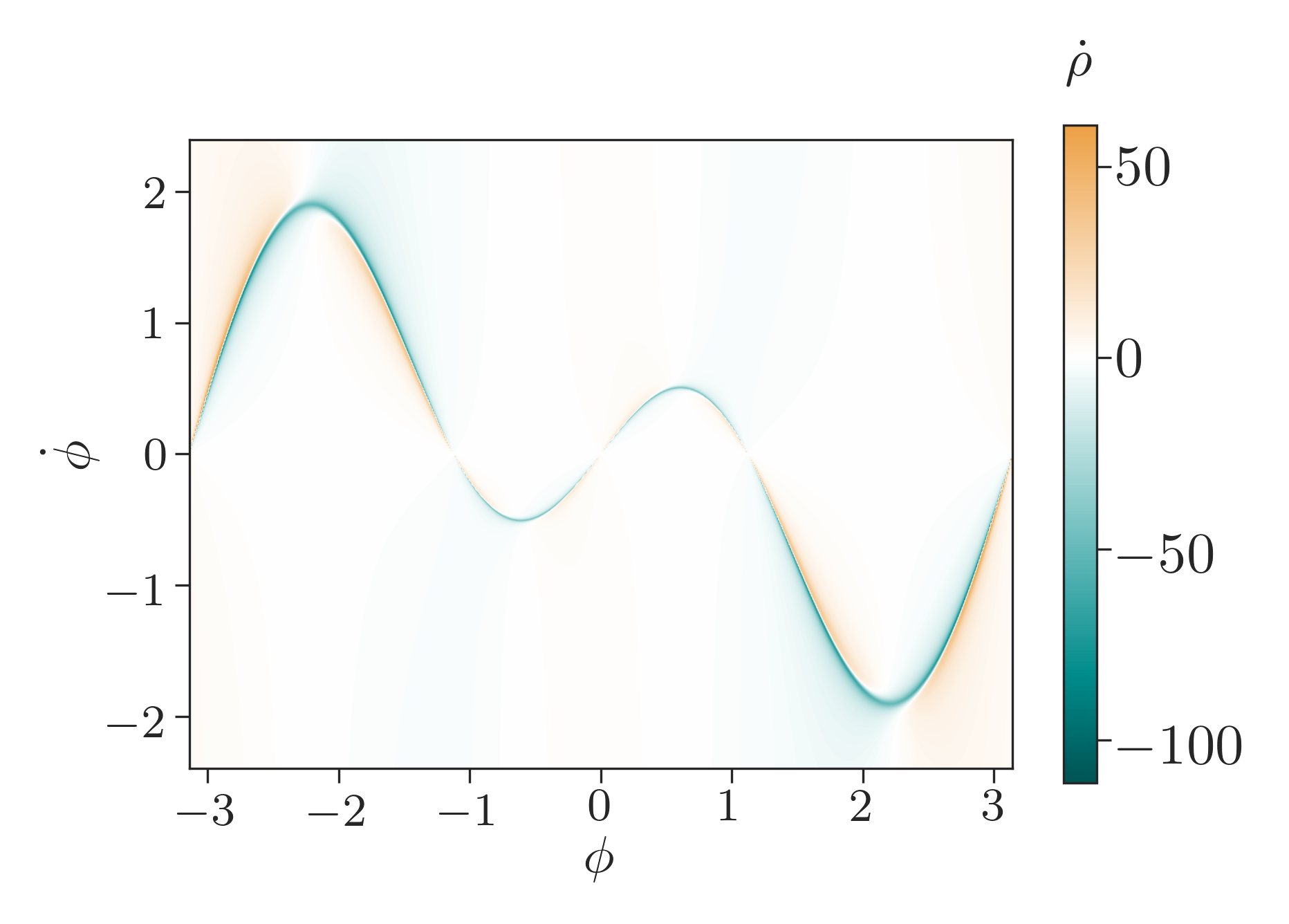}
\end{minipage}
\begin{minipage}{3.2in}
\centering
\includegraphics[width=3.2in]{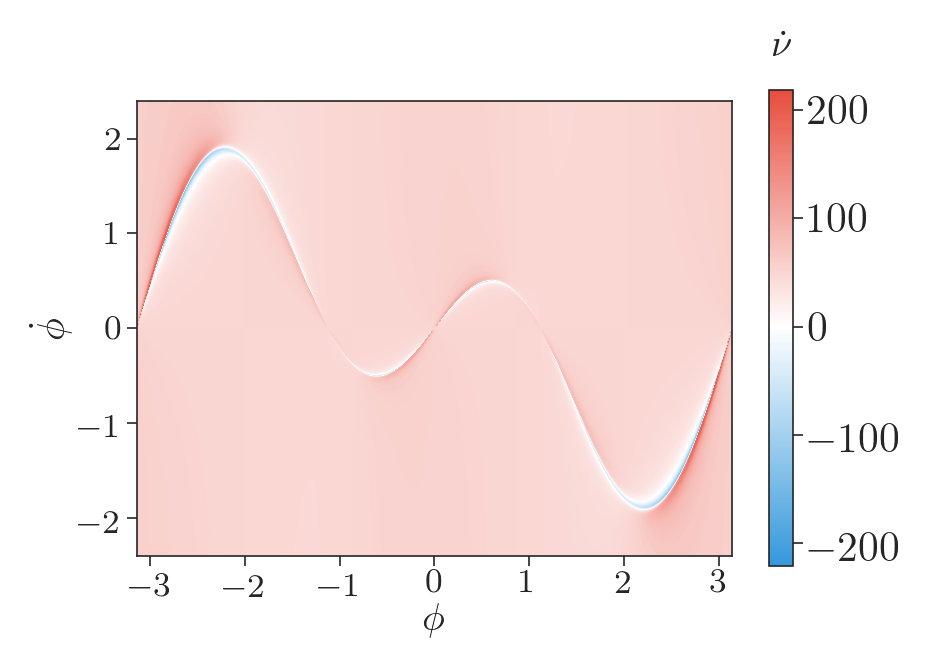}
\end{minipage}
\caption{\label{fig:rotHoopB} (top) The trajectory divergence rate and (bottom) the trajectory divergence ratio of Example \ref{ex:rotHoop} with $\varepsilon = 0.02$ and $\mu = 2.3$. These diagnostics both show strong attraction along the slow manifold visible above in Fig. \ref{fig:rotHoopA}.}
\end{figure}

For the parameters shown in Figs. \ref{fig:rotHoopA} and \ref{fig:rotHoopB}, both $\phi=0$, representing the bottom of the hoop, and $\phi=\pi$, representing the top of the hoop, give unstable fixed points, and a pair of fixed points on either side of the bottom of the hoop, at $\phi=\pm\arccos\tfrac{1}{\mu}$ become stable fixed points.

Figure \ref{fig:rotHoopB} shows that the trajectory divergence rate is quite effective at capturing the slow manifold for this example. The trough in the divergence rate field gives the attracting slow manifold onto which all trajectories converge. The trajectory divergence rate disappears near the equilibria of the system which are found each time the curve crosses the horizontal axis. As the trajectory divergence rate is calculated with normal vectors which are normalized by the magnitude of velocity, it is undefined at precisely the equilibria of the system.

\subsubsection{Search for the most attractive invariant manifold}
A trajectory which locally minimizes the trajectory divergence rate may be identified as the most attractive invariant manifold in phase space. The arc-length averaged trajectory divergence rate along a trajectory $\gamma$ gives a way to identify this minimizing trajectory, defined by,
\begin{equation}
\dot{P}=\frac{1}{\sigma(\gamma)}\left(\int_\gamma\dot{\rho}ds\right),
\end{equation}
where $\sigma(\gamma)=\int_{\gamma}ds$ is the arc length of $\gamma$.

Next, the arc-length averaged divergence rate $\dot{P}$ is minimized among a collection of trajectories $\Gamma$ to find a locally minimizing trajectory $\gamma^*$,
\begin{equation}\label{eq:minimum}
\gamma^*=\min_{\gamma\in\Gamma}\dot{P}.
\end{equation}
\begin{figure}
\centering
\includegraphics[width=3.2in]{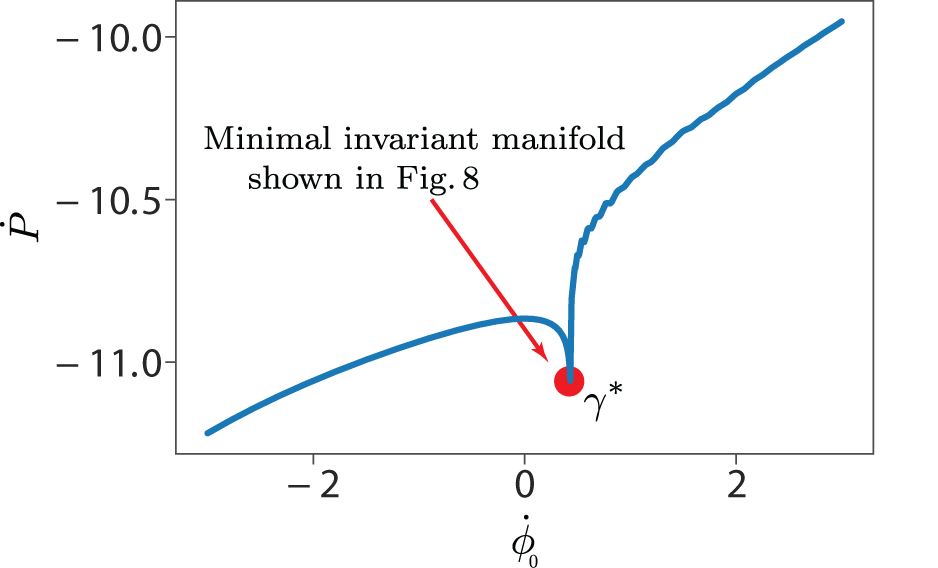}
\caption{The arc-length averaged trajectory divergence rate $\dot{\rho}$ over 1,000 trajectories integrated both forward and backward along the vertical slice $\phi=-3$.}
\label{fig:min-rhodot}
\end{figure}

\begin{figure}
	\centering
	\includegraphics[width=3.2in]{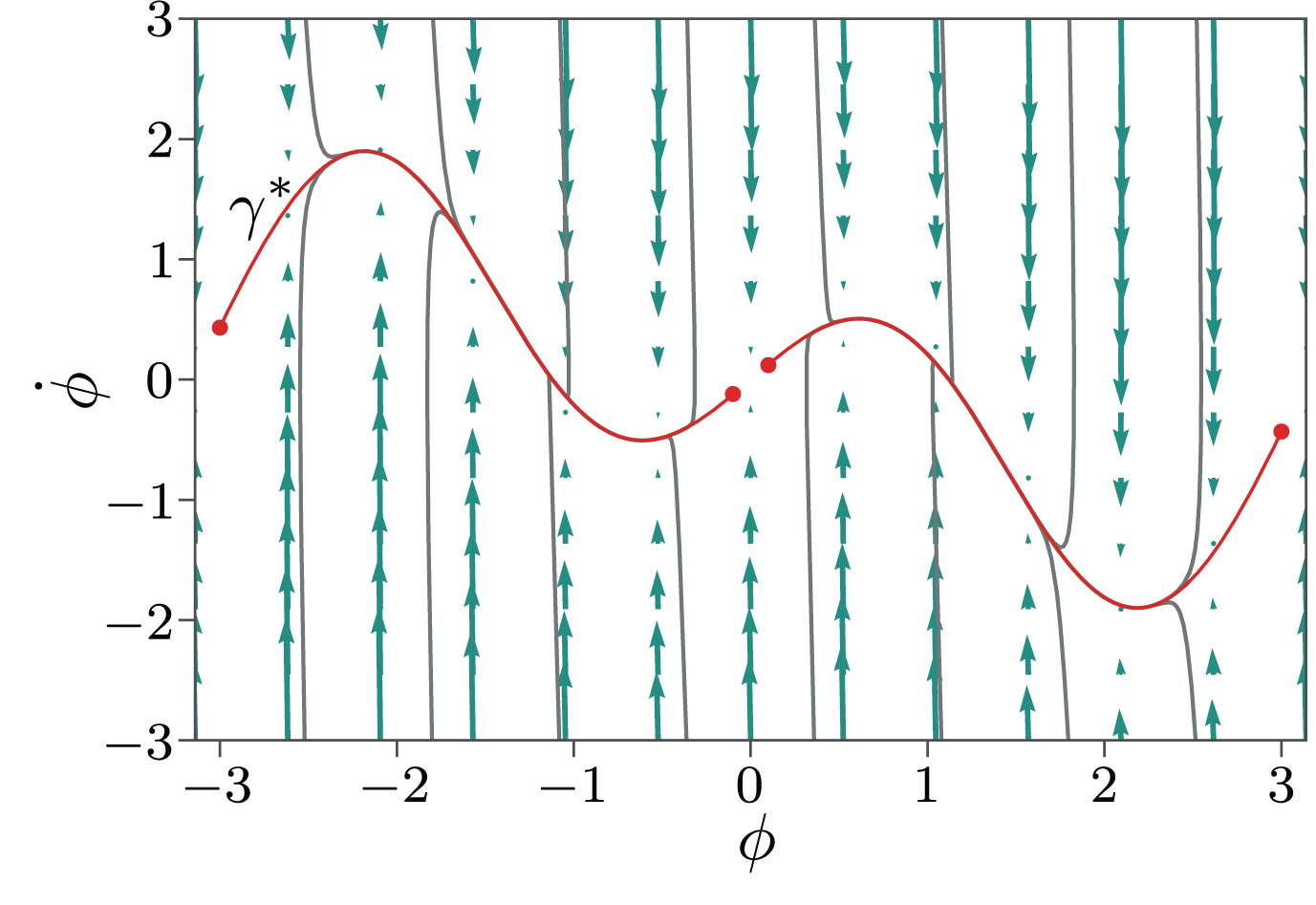}
	\caption{The results of the minimization of (\ref{eq:minimum}) as shown in Fig. \ref{fig:min-rhodot}. The red dot shows the local minimum point found from the simulations, and the red curve shows the forward integration of this trajectory. Local minima were calculated at the points $\phi=\{-3.0, -0.1, 0.1, 3.0\}$ and integrated forward.}
	\label{fig:min-rhodot-comp}
\end{figure}

Fig. \ref{fig:min-rhodot} shows the result of the calculation of arc-length averaged divergence rate for trajectories intersecting $\phi=-3$ over a range $\dot{\phi}_0\in[-3, 3]$. These trajectories were integrated forward until they reached a fixed point and integrated backward until they reached $\dot{\phi}=\pm11$. The results of this numerical minimization are shown in Fig. \ref{fig:min-rhodot-comp} by identifying the locally minimizing trajectory $\gamma^*$ by its intersection with $\phi=-3$ by the red circle and showing its forward integration in red. This process is repeated for $\phi=-0.1$, and by symmetry, the values are obtained for $\phi=0.1$ and $\phi=3.0$.

\subsubsection{Comparison with other methods}
As an example of a situation where trajectory-restricted measurements are useful, consider the following example from \cite{haller_variational_2011}, presented in the introduction of the trajectory-normal repulsion rate.

\begin{figure}
\centering
\includegraphics[width=3.1in]{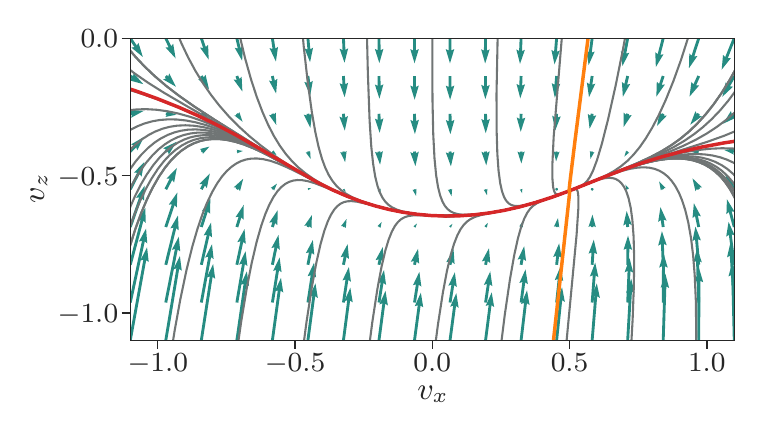}
\caption{The phase portrait for Eq. (\ref{eq:glider}) in Example \ref{ex:glider} with $\theta=-5^\circ$, showing the weak stable manifold (red) and strong stable manifold (orange) passing through the stable node fixed point at (0.50, -0.56).}
\label{fig:glider}
\end{figure}

\example{\it\textendash \label{ex:glider} The terminal velocity manifold in gliding flight.} \\
In a simplified model of passive gliding flight, a globally attracting codimension-one manifold may be observed in the glider's velocity space \cite{nave2018global,yeaton_global_2017}. Because every trajectory is rapidly attracted to this structure and evolves along or near it, it serves as a higher dimensional analogue to terminal velocity and is therefore referred to as the terminal velocity manifold. The glider's motion under this model is given by the nondimensional equations of motion,
\begin{equation}
\begin{aligned}
\dot{v}_x &= \left(v_x^2+v_z^2\right)\left(C_L\left(\gamma+\theta\right)\sin\gamma - C_D\left(\gamma+\theta\right)\cos\gamma\right), \\
\dot{v}_z &= \left(v_x^2+v_z^2\right)\left(C_L\left(\gamma+\theta\right)\cos\gamma + C_D\left(\gamma+\theta\right)\sin\gamma\right) - 1,
\end{aligned}
\label{eq:glider}
\end{equation}
where $v_x=\tfrac{\epsilon}{gc}V_x$ and $v_z=\tfrac{\epsilon}{gc}V_z$ are the dimensionless horizontal and vertical components of velocity, $\theta$ is the body's fixed pitch angle with respect to the ground, $\gamma=\arctan\tfrac{-v_z}{v_x}$ is the angular direction of motion of the body, $C_L$ and $C_D$ represent lift and drag as functions of angle of attack $\gamma+\theta$, and derivatives are with respect to dimensionless time $t=\sqrt{c\epsilon/g}T$. The universal glide scaling parameter $\epsilon=\tfrac{\rho c S}{2m}$ included in these scalings allows for comparison of different gliders with the same equations \cite{yeaton_global_2017}. Within these dimensionless variables are the horizontal and vertical velocity $V_x,V_z$, chord length $c$, glider span $S$, gravity $g$, time $T$, fluid density $\rho$, and mass $m$.

The phase portrait of this example for a falling flat plate with a fixed pitch of $\theta=-5^\circ$ is shown in Fig. \ref{fig:glider}. From the stable fixed point located at $(v_x,v_z)=(0.50, -0.56)$, there are both strong and weak stable submanifolds within the 2-dimensional stable manifold. Because many invariant manifolds intersect the origin with the same tangent direction, the weak stable submanifold is nonunique, and methods such as the trajectory-normal repulsion rate may be used to identify the most influential weak stable submanifold \cite{haller_variational_2011,nave2018global}.

\begin{figure*}
\centering
\includegraphics[width=\linewidth]{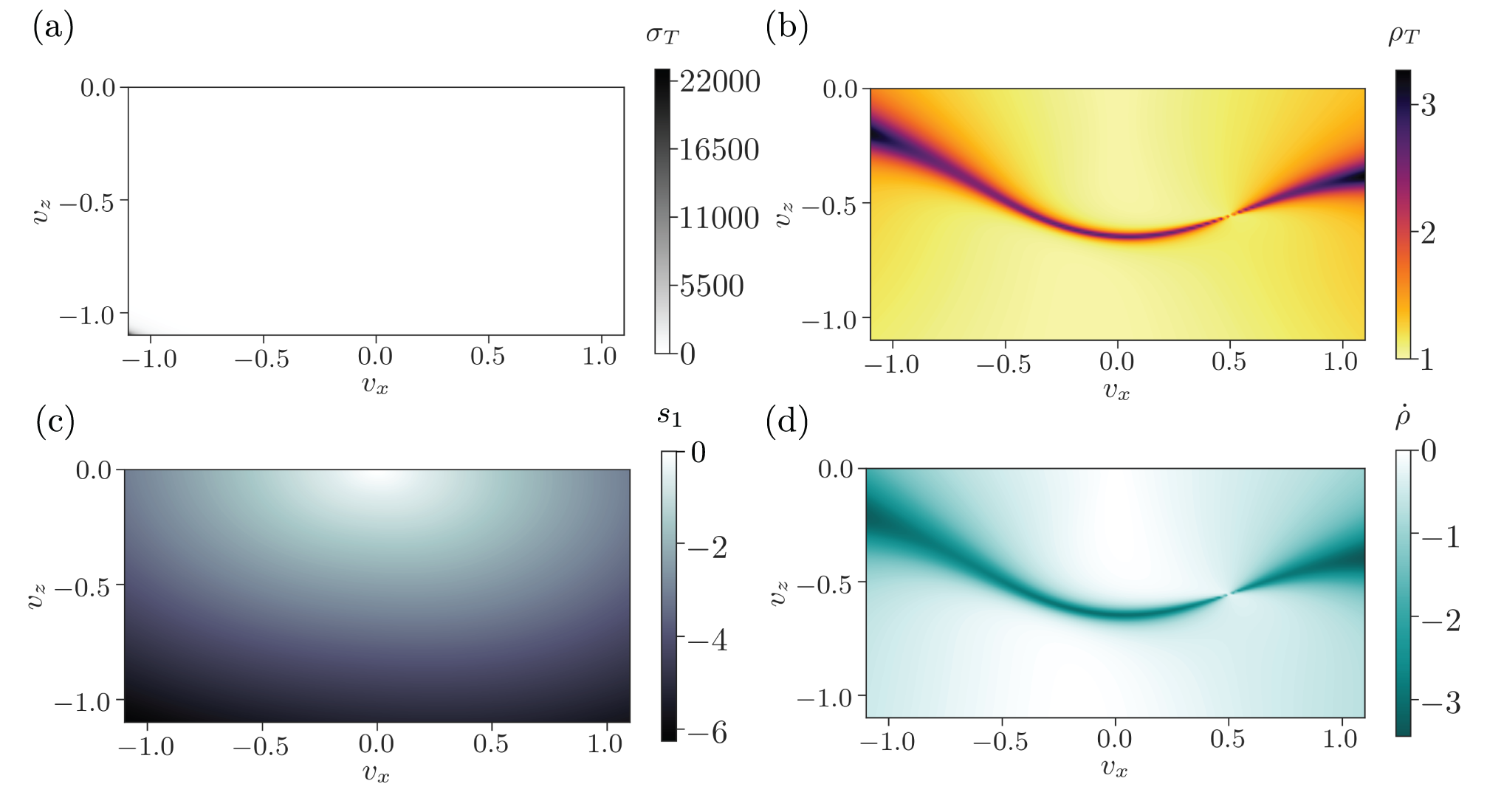}
\caption{Comparison of methods to extract structure from Ex. \ref{ex:glider}. The scalar field calculations using (a) the finite-time Lyapunov exponent \cite{shadden2005definition}, (b) the trajectory-normal repulsion rate \cite{haller_variational_2011}, (c) minimum eigenvalue of $\mathbf{S}$ from the objective Eulerian coherent structure approach \cite{serra_objective_2016}, and (d) the trajectory divergence rate. Panels (a) and (b) were calculated by integrating trajectories over an integration time $T=-0.33$.}
\label{fig:method-comparison}
\end{figure*}

For this example, Fig. \ref{fig:method-comparison} presents the finite-time Lyapunov exponent $\sigma_T$, the trajectory-normal repulsion rate $\rho_T$, the minimum eigenvalue of the rate-of-strain tensor $\mathbf{S}$ denoted by $s_1$, and the trajectory divergence rate $\dot{\rho}$ over the domain as a comparison of these methods. The finite-time Lyapunov exponent $\sigma_T$ and eigenvalue of the rate-of-strain tensor $s_1$ both represent measures of pure stretching, and are therefore dominated by the tangential stretching. These measures both give no structure in the system. Hyperbolic objective Eulerian structures, as would be measured by $s_1$, must contain an isolated local maximum or minimum of $s_1$ \cite{serra_objective_2016}, and no such isolated maximum or minimum exists in this example. 

The two trajectory-based measures, on the other hand, both show the structure of the system, with the integrated measure of $\rho_T$ giving a more defined ridge by taking into account longer time information.

\subsection{Approximation of hyperbolic Lagrangian coherent structures}\label{ss:LCSapprox}
As discussed in the introduction, in fluid flows, it can be very useful to look at finite-time barriers to transport in the fluid, known as Lagrangian coherent structures (LCSs) \cite{shadden2005definition}.

Two key limitations of many Lagrangian methods for detecting flow structures are the computation time for advecting trajectories \cite{ameli2014development,brunton2010fast} and dealing with limited experimental data \cite{garth2007efficient}. By considering an Eulerian (i.e., instantaneous) approximation of Lagrangian structures, the trajectory divergence rate can provide a first look at the structure of a given vector field. It may even be applied to nonautonomous flows to show the attraction and repulsion of the vector field at each time step. Although there may be exceptions \cite{tallapragada2017globally}, in most cases the short time transport barriers of a system are locally attracting or repelling. Therefore, the trajectory divergence rate may be used to approximate Lagrangian coherent structures. 

\begin{figure*}
\centering
\includegraphics[width=\linewidth]{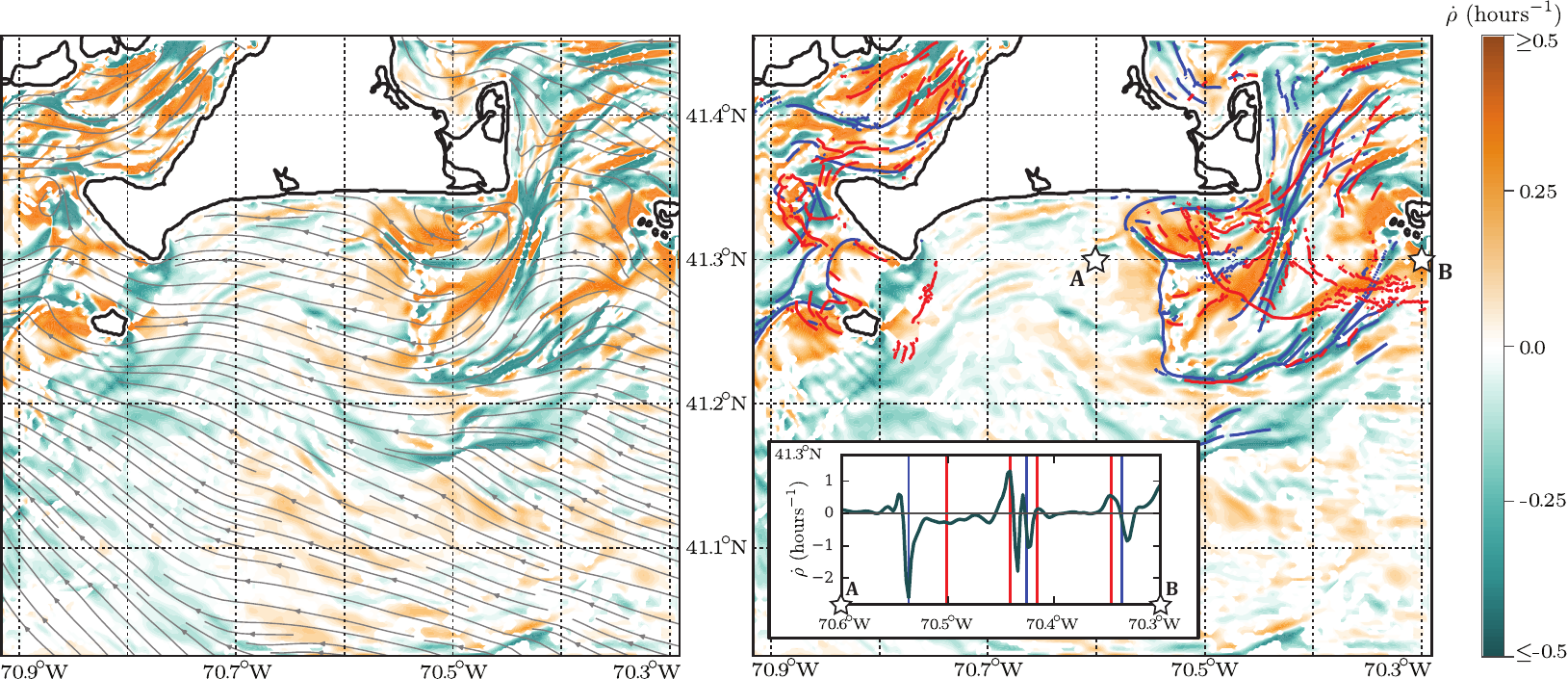}
\caption{Simulation of ocean flow around Martha's Vineyard on Monday, August 14, 2017 at 10:00AM EDT using the MSEAS model \cite{de2014relocatable}. Streamlines of the ocean flow are shown in gray in the left figure. Lagrangian coherent structures calculated using $C$-ridges \cite{schindler2012ridge} for an integration time of 2 hours are shown as solid lines in the right figure, with attracting LCSs in blue and repelling LCSs in red. The trajectory divergence rate, shown in the background shows regions of attraction (cyan) and repulsion (orange). The inset figure shows the trajectory divergence rate $\dot{\rho}$ along $41.3^\circ$N latitude from $70.6^\circ$W to $70.3^\circ$W, marked by the points A and B. Intersecting attracting and repelling LCSs are shown with the blue and red vertical lines, respectively.}
\label{fig:LCScomp}
\end{figure*}

\example{\it\textendash Data-driven ocean model.} \\
This example presents simulations in Fig. \ref{fig:LCScomp} of the ocean flow around Martha's Vineyard, Massachusetts in August 2017 using the MIT-MSEAS primitive-equation ocean model \cite{haley2010multiscale}. With this simulation, it is possible to calculate trajectories of advected ocean flow to generate the flow map and therefore the right Cauchy Green tensor. The Lagrangian coherent structures are identified by calculating derivatives of the finite-time Lyapunov exponent in the largest eigendirection of the right Cauchy-Green tensor to find $C$-ridges \cite{schindler2012ridge}. These quantities are calculated over a forward or backward integration time of 2 hours for repelling and attracting features, respectively. In Fig. \ref{fig:LCScomp}, it is clear that many of the LCSs align with regions of high repulsion or attraction and generally follow the ridges of these highly attracting or repulsive regions.

The trajectory divergence rate, because it measures normal growth of normal vectors, will be most aligned with the finite-time Lyapunov exponent, which measures maximum growth, when the direction of maximum growth occurs normal to a trajectory. In addition, because FTLE is a time-integrated measure, LCS will be most easily approximated by the trajectory divergence rate when the vector field is changing slowly relative to the dynamics.

Ridges and valleys of the trajectory divergence rate do not exactly correspond with hyperbolic LCSs, but they can serve as an approximation for a much lower computational cost. The trajectory divergence rate, for large geophysical fluid flows, can prove particularly helpful for identifying regions of interest in the flow quickly. This can be very important in search-and-rescue situations, or may be used as the first step in, for instance, an adaptive mesh algorithm to identify finite-time structures \cite{lekien2010computation,xie2018lagrangian}.

\subsection{Repulsion rate along a limit cycle}
Because the trajectory divergence rate $\dot{\rho}$ gives the instantaneous attraction or repulsion of nearby trajectories, it can be applied to analyze the local stability of trajectories. For instance, it has been shown that globally attracting limit cycles may be locally repelling in places, which has significant implications for control \cite{ali1999local,norris2008revisiting}. As an example, consider the classical Van der Pol oscillator.

\begin{figure}
\centering
\includegraphics[width=2.8in]{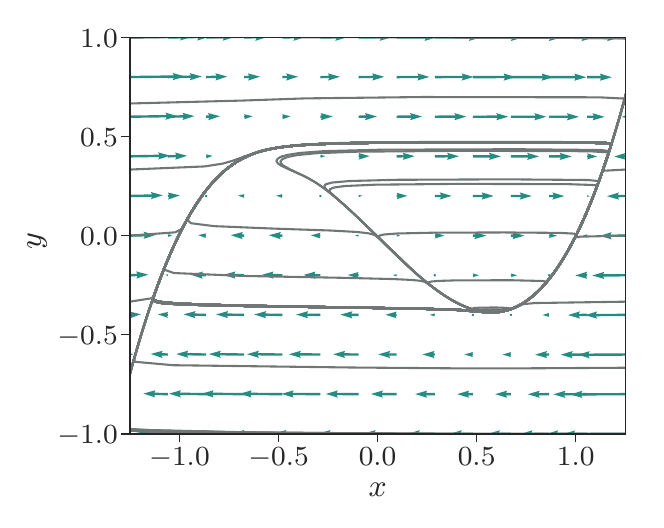}
\caption{The phase portrait of the Van der Pol oscillator given by Example \ref{ex:vdP}, using $\varepsilon = 0.01$ and $a = 0.575$.}
\label{fig:phaseplot-vdP}
\end{figure}

\example{\it\textendash \label{ex:vdP} Van der Pol oscillator.} \\
The Van der Pol oscillator is another slow-fast system, but the folds along the slow manifold separate stable and unstable branches, which admit a so-called ``canard explosion,'' leading to a limit cycle \cite{krupa2001relaxation,kuehn2016multiple}. The limit cycle is globally attracting, with two branches evolving along the slow manifold and the other two moving quickly across the system, as shown in Fig. \ref{fig:phaseplot-vdP}.

The governing equations of the system are given by,
\begin{equation}
\begin{aligned}
\dot{x} & = \frac{1}{\varepsilon} \left(y + x - x^3\right), \\
\dot{y} & = a - x,
\end{aligned}
\label{eq: vanderPol}
\end{equation}
with $0<\varepsilon\ll1$ and $a\in\mathbb{R}$. In this system, the slow dynamics occur along a slow manifold near $y=-x+x^3$, which is the critical manifold of the system. As the parameter $a$ is increased from $0$, the location of the fixed point changes until a Hopf bifurcation occurs, first forming a small limit cycle around the fold of the slow manifold before expanding to a loop around both fold points.

\begin{figure*}
\centering
\includegraphics[width=0.49\textwidth]{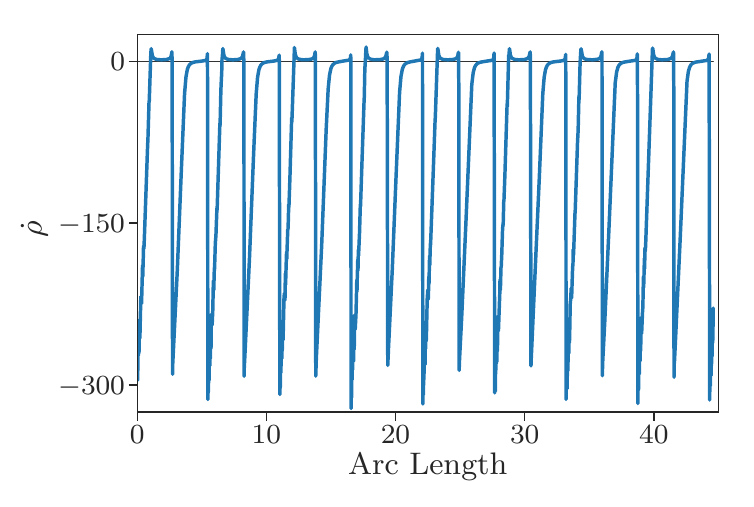}
\includegraphics[width=0.49\textwidth]{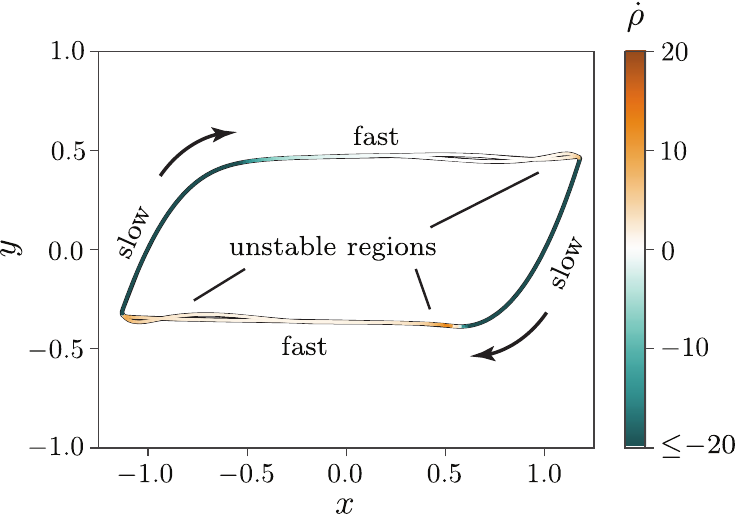}
\caption{The trajectory divergence rate along the limit cycle in the Van der Pol oscillator of Example \ref{ex:vdP}. Along the limit cycle, the outer branches along the slow manifold are very attractive, but the fast branches are instantaneously repelling.}
\label{fig:limitcycle}
\end{figure*}

Several works have shown that although a limit cycle may be globally stable, it is not always locally stable \cite{ali1999local,norris2008revisiting}. That is, there may be regions of a globally attracting limit cycle which are locally repelling. These regions may play a critical role in attempts to control dynamics which naturally occur in an oscillatory way, identifying the location along the limit cycle trajectory most sensitive to perturbation. The trajectory divergence rate along the trajectory, shown in Figure \ref{fig:limitcycle}, provides an excellent tool for looking at this sensitivity to perturbation along a limit cycle. In this example it is applied to the Van der Pol oscillator of Example \ref{ex:vdP}. The cycle is attracting for most of its space, but there are moments, when moving across the fold point to the opposite stable branch, when the trajectory is normally repelling.

\section{Extension to higher dimensions}\label{s:HigherDimension}
With the intuition of the trajectory divergence rate in two dimensions, this method may be extended to higher dimensions. While the derivation used in Section \ref{s:Derivation} relies on the 2-dimensionality of the system, an understanding of the results of the derivation allows for an extension of the concept to higher dimensions. The two dimensional trajectory divergence rate calculates the normal projection of the instantaneous rate of deformation of the unique trajectory-normal vector $\dot{\rho} = \langle \mathbf{n}, \mathbf{S}\mathbf{n}\rangle$. In higher dimensions, trajectories remain one dimensional, so the normal direction becomes a normal hyperplane. Therefore, in higher dimensions, the normal projection of the instantaneous rate of deformation of the normal hyperplane, $\mathbf{N}$, may be written using the analogous formula,
\begin{equation}\label{eq:higherdim-reprate}
\dot{\mathbf{R}} = \mathbf{N}^\dagger \mathbf{S}\mathbf{N}.
\end{equation}
In $\mathbb{R}^k$, $\mathbf{S}$ is the $k\times k$ rate-of-strain tensor, still given by $\mathbf{S}=\tfrac{1}{2}\left(\nabla \mathbf{v}+\nabla \mathbf{v}^\dagger\right)$ and $\mathbf{N}$ is the $k \times (k-1)$ matrix representing the hyperplane normal to the tangent vector $\mathbf{v}$. Therefore $\dot{R}$ will be of dimension $(k-1) \times (k-1)$. The eigenvalues of $\dot{R}$ give the principal trajectory divergence rates and the eigenvectors give the directions. The maximal eigenvalue of $\dot{R}$ gives the dominant attraction or repulsion of every point. The lower eigenvalues give the dimension of this attraction.

\begin{figure*}
\centering
\includegraphics[width=5in]{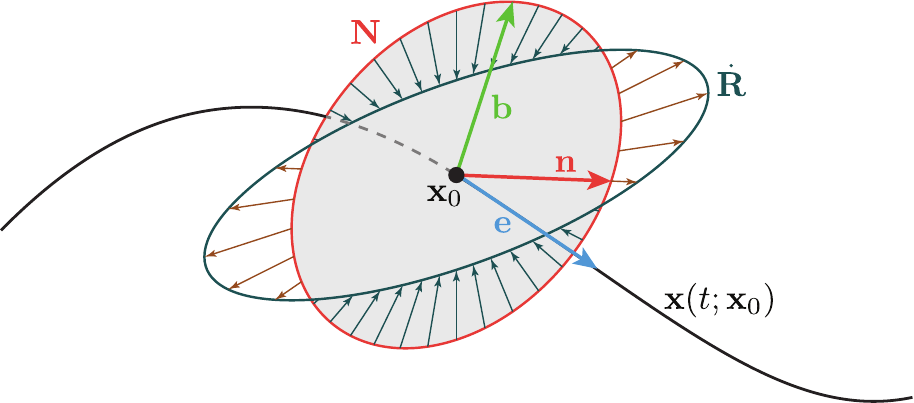}
\caption{Representation of the divergence rate in a 3-dimensional system, $\dot{\textbf{R}}$. The circle represents the normal plane $N$ to the trajectory, and the ellipse represents the slope of normal stretching as in Fig. \ref{fig:div-rate-schematic}. The eigenvalues of $\dot{\mathbf{R}}$ give the principal stretching within the normal plane.}
\label{fig:3dDivRateSchematic}
\end{figure*}

\begin{figure*}
	\centering
	\includegraphics[width=0.85\linewidth]{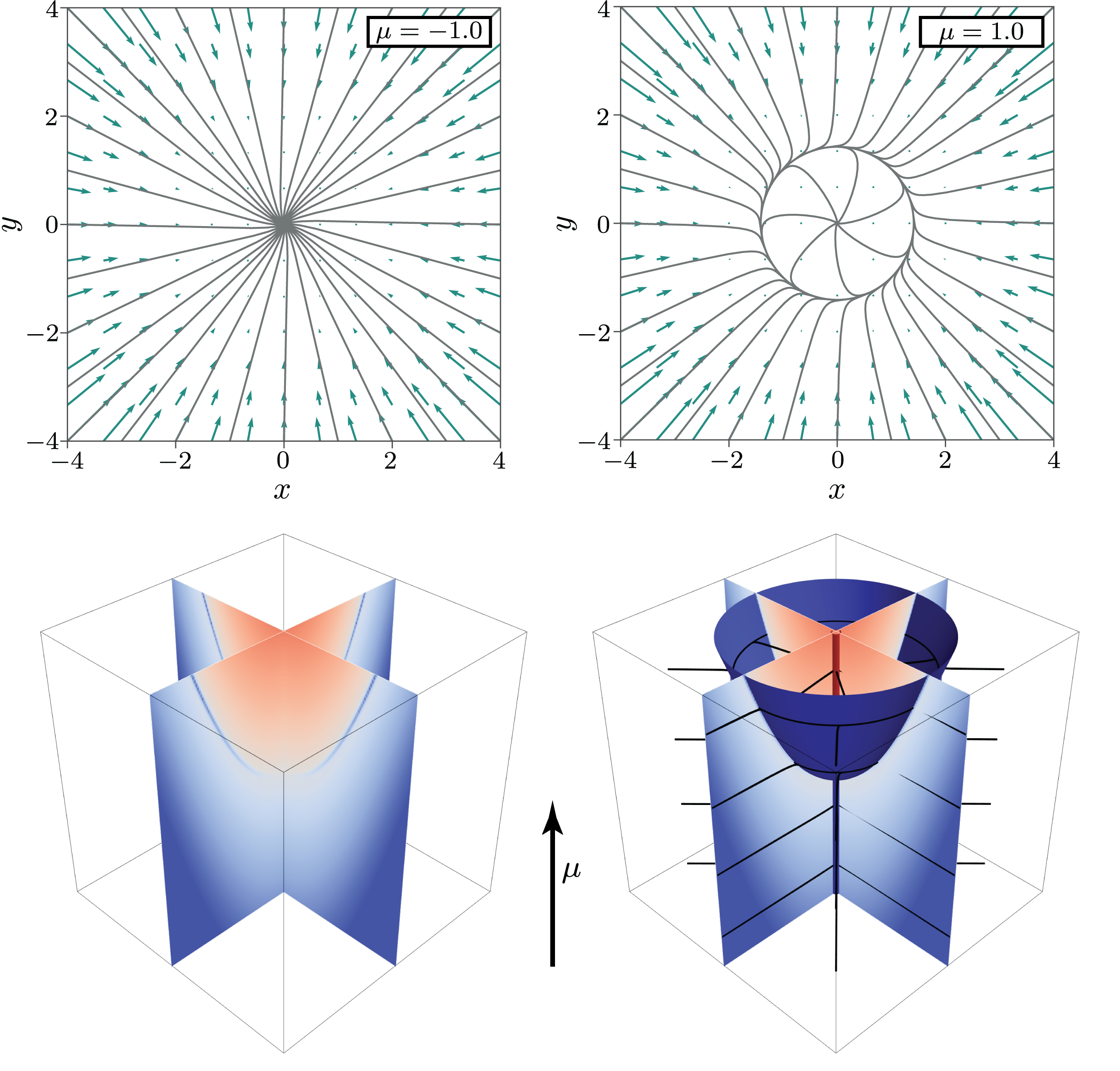}
	\caption{The phase portrait for Example \ref{ex:hopf} with $\mu=-1$ and $\mu=1$, and results of the 3D computation of the trajectory divergence rate. In the bottom left, two slices of the largest eigenvalue of $\dot{\mathbf{R}}$ from Eq. (\ref{eq:higherdim-reprate}) are shown for the Hopf bifurcation in extended phase space as described by (\ref{eq:Hopf}), with $\epsilon=0.25$. In the bottom right, this has been overlaid with the attracting manifold (blue) and example trajectories.}
	\label{fig:Hopf}
\end{figure*}

In 3 dimensions, the Frenet-Serret reference frame gives the normal plane to a trajectory $\gamma$ with a tangent vector $\gamma'$ and an acceleration $\gamma''$. In this frame, the normal plane is defined by two vectors: the normal $\mathbf{n}$ and binormal $\mathbf{b}$. The normal vector is defined to be in the direction of change of the tangent vector $\mathbf{e}$.
\begin{equation}
\begin{aligned}
{\mathbf{n}} &= \frac{\tfrac{d{\mathbf{e}}}{ds}}{\left|\tfrac{d{\mathbf{e}}}{ds}\right|}, \\
\frac{d{\mathbf{e}}}{ds} &= \frac{\gamma'\times\gamma''}{\left|\gamma'\right|^3}.
\end{aligned}
\end{equation}
The binormal, which must be perpendicular to both the tangent and normal vectors, can be found with their cross product,
\begin{equation}
{\mathbf{b}} = {\mathbf{e}}\times{\mathbf{n}}.
\end{equation}
The normal plane $\mathbf{N}$ is then defined by the normal and binormal unit vectors,
\begin{equation}
\mathbf{N} = \left[{\mathbf{n}},\,{\mathbf{b}}\right].
\end{equation}
With this definition of the normal plane in 3 dimensions, (\ref{eq:higherdim-reprate}) may be rewritten as,
\begin{equation}
\dot{R} = \left[\begin{array}{cc}
\langle{\mathbf{n}}, \mathbf{S}{\mathbf{n}}\rangle & \langle{\mathbf{n}}, \mathbf{S}{\mathbf{b}}\rangle \\
\langle{\mathbf{b}}, \mathbf{S}{\mathbf{n}}\rangle & \langle{\mathbf{b}}, \mathbf{S}{\mathbf{b}}\rangle  \\
\end{array}\right].
\end{equation}
Calculating the higher-dimensional trajectory divergence rate of (\ref{eq:higherdim-reprate}) shows the rates of stretching of the plane normal to each trajectory. Fig. \ref{fig:3dDivRateSchematic} shows a visualization of this interpretation. There are an infinite number of vectors normal to the trajectory, so the higher-dimensional trajectory divergence rate $\dot{\mathbf{R}}$ measures the rates of deformation of the entire plane, and eigenvalues of this matrix give the principal magnitudes and directions of repulsion and attraction.

\example{\it\textendash Supercritical Hopf bifurcation \label{ex:hopf}.} \\
The Poincar\'e-Andronov-Hopf, or simply Hopf, bifurcation is one of the most fundamental to nonlinear dynamics. As a parameter is increased, a single fixed point reverses its stability as a limit cycle appears. In parameter extended phase space, the Hopf bifurcation becomes a three-dimensional system, with the bifurcation parameter $\mu$ representing the third dimension \cite{wiggins2003introduction}, in which case the limit cycle is represented as a paraboloid. As an example of the application of the 3-dimensional trajectory divergence rate $\dot{\textbf{R}}$ from Eq. (\ref{eq:higherdim-reprate}), consider a slow-fast version of the Hopf normal form, with attraction to the limit cycle moving at a faster time scale than motion around the limit cycle. The parameter dynamics remain trivial.	
\begin{equation}
\begin{aligned}
\dot{x} &= \frac{1}{\epsilon}\left(2\mu-\left(x^2+y^2\right)\right)x-y, \\
\dot{y} &= \frac{1}{\epsilon}\left(2\mu-\left(x^2+y^2\right)\right)y+x, \\
\dot{\mu} &= 0.
\end{aligned}
\label{eq:Hopf}
\end{equation}
For a supercritical Hopf bifurcation, $\mu<0$ corresponds with a single stable fixed point at the origin and $\mu>0$ corresponds to an unstable fixed point at the origin with an attracting limit cycle of radius $\sqrt{2\mu}$.

Fig. \ref{fig:Hopf} shows the phase space for this example for both $\mu=-1$, below the bifurcation, and $\mu=1$, beyond the bifurcation point, showing the appearance of the limit cycle. Calculating the largest magnitude eigenvalue of the 3-dimensional repulsion rate $\dot{\mathbf{R}}$ shows the dominant attraction or repulsion at each point in extended parameter space. The panel shows the values along the $(x,\mu)$ and $(y,\mu)$ planes, to represent the 3-dimensional data. Particularly of note is the narrow dark blue region in the top half of this panel, indicating the attracting limit cycle of the system. This method is unable to identify the attracting line which exists for $\mu<0$, but is able to calculate the attracting paraboloid, and shows instability in the center of the paraboloid, with a peak closest to the center. This example shows promise for the future application of the trajectory divergence rate in higher dimensions.

\section{Summary and conclusions}\label{s:Summary}	
The trajectory divergence rate is an inherent property of continuously differentiable vector fields that naturally follows from either the instantaneous stretching of vectors or the trajectory-normal repulsion rate. It measures the rate at which the normal distance between nearby trajectories grows at every position in the domain of the vector field. It is a straightforward quantity to compute, requiring only the instantaneous vector field and its gradient, and therefore may provide a useful diagnostic when investigating the geometric properties of a flow. As shown in the case of the finite-time Lyapunov exponent \cite{lekien2010computation}, gradients are also computable on unstructured meshes.

In application, the trajectory divergence rate and divergence ratio may be applied to approximate slow manifolds, weak stable or unstable manifolds, or hyperbolic Lagrangian coherent structures, or to measure the local stability of trajectories such as limit cycles.

The trajectory divergence rate and ratio are computationally efficient and physically intuitive. These scalar fields may become useful tools for the investigation of many applications in various fields. The python package ManifoldID for manifold identification, developed for this paper, may be found on GitHub at  \href{https://github.com/gknave/manifoldid}{https://github.com/gknave/manifoldid}.

\section*{Conflict of interest}
The authors declare that they have no conflict of interest.

\bibliographystyle{spmpsci}
\bibliography{div-rate-nave-ross}

\appendix

\section{Derivation of Equation (\ref{eq:DivRate})} \label{ap: normal derivation}
Starting with (\ref{eq:local Rho}), the trajectory-normal repulsion rate, \(\rho_T\) can be written, to leading order in $T$, as,
\begin{equation}
\begin{aligned}
\rho_T &= 1+ \left(\text{tr}(\mathbf{S})-\frac{\mathbf{v}^\dagger\mathbf{S}\mathbf{v}}{\left|\mathbf{v}\right|^2}\right)T \\
&= 1+ \frac{1}{\left|\mathbf{v}\right|^2}(\text{tr}(\mathbf{S})\left|\mathbf{v}\right|^2-\mathbf{v}^\dagger\mathbf{S}\mathbf{v})T  \\
&= 1+ \frac{\mathbf{v}^\dagger(\text{tr}(\mathbf{S})\mathbf{I} - \mathbf{S})\mathbf{v}}{\left|\mathbf{v}\right|^2}T
\end{aligned}
\end{equation}

For a 2-tensor, \(\mathbf{A}\), the relation \(\text{tr}(\mathbf{A})\mathbf{I}-\mathbf{A}\) in index notation may be written as \(A_{ii}\delta_{jk} - A_{jk}\).
\begin{equation}
\begin{aligned}
\text{tr}(\mathbf{A})\mathbf{I}-\mathbf{A} &= A_{ii}\delta_{jk} - A_{jk} \\
&= A_{il}\delta_{li}\delta_{jk} - A_{il}\delta_{lk}\delta_{ji} \\
&= A_{il}(\delta_{li}\delta_{jk} - \delta_{lk}\delta_{ji}) \\ 
&= A_{il}\varepsilon_{lj}\varepsilon_{ik} \\
&= \mathbf{R}^\dagger\mathbf{A}\mathbf{R}
\end{aligned}
\end{equation}
Where \(\varepsilon_{ij}\) is the 2-dimensional Levi-Cevita symbol which, for a 2x2 matrix, is the index representation of the negative of the $90^\circ$ counter-clockwise rotation matrix,  $\varepsilon_{ij} = -\mathbf{R}$. Therefore, for small time $T$, \(\rho_T\) may be written as,
\begin{equation}
\begin{aligned}
\rho_T &= 1+ \frac{\mathbf{v}^\dagger(\mathbf{R}^\dagger\mathbf{S}\mathbf{R})\mathbf{v}}{\left|\mathbf{v}\right|^2}T \\
& = 1+ \frac{(\mathbf{R}\mathbf{v})^\dagger\mathbf{S}(\mathbf{R}\mathbf{v})}{\left|\mathbf{v}\right|^2}T 
\end{aligned}
\end{equation}
Which can  alternatively be written in terms of the unit normal field, \(\mathbf{n} = \mathbf{R}\mathbf{v}/\left|\mathbf{v}\right|\), as in (\ref{unit_vector_fields}), yielding
\begin{equation}
\rho_T 
= 1+ \langle \mathbf{n},\mathbf{S}\mathbf{n} \rangle T 
\end{equation}	
which gives the leading order behavior defined by the instantaneous rate,
\begin{equation}	
\dot \rho = \langle \mathbf{n},\mathbf{S}\mathbf{n} \rangle
\end{equation}
Note that the rate of length change for an infinitesimal material element vector $\ell$ based at $\mathbf{x}_0$ and advected under the flow is
\begin{equation}
\frac{d}{dt} |\ell| = \frac{1}{| \ell |} \langle \ell,\mathbf{S}\ell \rangle
\end{equation}
Thus, the leading order behavior of the trajectory-normal repulsion rate for short time \(T\) can be thought of as the rate  of stretching of unit normal vectors, normal to the invariant manifold passing through $\mathbf{x}_0$.
This value is locally maximized along the most repulsive (or attractive) manifolds, which provide the most influential core of phase space deformation patterns.

\end{document}